\newcommand{\printstyle}{reprint}
\newcommand{\halfwidth}{\columnwidth} % use with printstyle = reprint (replicates Journal layout) 

\documentclass[aip,pop,floatfix,\printstyle ,nofootinbib]{revtex4-1}

\usepackage{color}
\usepackage[caption=false]{subfig}
\usepackage{graphicx,booktabs,array}
\usepackage{multirow}
\DeclareGraphicsExtensions{.pdf,.png,.jpg,.eps}
\usepackage{epsfig}
\usepackage{epstopdf}
\usepackage{amssymb}
\usepackage{amsmath}
\usepackage{amsfonts}
\usepackage{mathrsfs}
\usepackage{amsthm}
\usepackage{float}
\usepackage{amsbsy}
\usepackage{bm}
\usepackage{grffile}
\usepackage{hyperref}
\hypersetup{
    colorlinks = true,
    citecolor = blue,
    urlcolor = blue,
    linkcolor = blue,
}
\usepackage{courier}
\usepackage{upgreek}
\usepackage{xspace}
\usepackage{xcolor}
\newcommand{\eg}{\textit{e}.\textit{g}. }

\newcommand{\Alfven}{Alfv\'{e}n }
\newcommand{\Alfvenic}{Alfv\'{e}nic }

\newcommand{\va}{v_A}

\newcommand{\vb}{v_0}

\newcommand{\omegaci}{\omega_{ci}}
\newcommand{\omegacio}{\omega_{ci0}} % used for on-axis cyclotron frequency 

\newcommand{\omeganorm}{\omega / \omegaci}

\newcommand{\omegabar}{\bar{\omega}}
\newcommand{\omegace}{\omega_{ce}}
\newcommand{\omegape}{\omega_{pe}}

\newcommand{\db}{\delta B}

\newcommand{\dbpar}{\db_\parallel}

\newcommand{\kpar}{k_\parallel}
\newcommand{\kperp}{k_\perp}

\newcommand{\kperpmin}{k_{\perp,\text{min}}}

\newcommand{\pphi}{p_\phi}

\newcommand{\vinj}{\vb/\va}

\newcommand{\linj}{\lambda_0}
\newcommand{\dl}{\Delta\lambda}

\newcommand{\vpar}{v_\parallel}
\newcommand{\vpres}{v_{\parallel,\text{res}}}

\newcommand{\vdrift}{v_{\text{Dr}}}
\newcommand{\vperp}{v_\perp}
\newcommand{\vperpz}{v_{\perp,0}}
\newcommand{\lres}{\ell}

\newcommand{\J}[2]{\mathscr{J}_{#1}^{#2}}
\newcommand{\Jlm}{\J{\lres}{m}}

\newcommand{\Jzm}{\J{0}{m}}
\newcommand{\Jzg}{\J{0}{G}}
\newcommand{\Jzc}{\J{0}{C}}

\newcommand{\Bres}{\eta}
\newcommand{\bres}{\Bres}

\newcommand{\W}{\mathcal{E}}

\newcommand{\vk}{\vec{k}}
\newcommand{\vdB}{\delta\vec{B}}
\newcommand{\vkperp}{\vec{k}_\perp}

\newcommand{\xinj}{x_0}
\newcommand{\dx}{\Delta x}

\newcommand{\fb}{f_0}

\newcommand{\omegacires}{\avg{\bar{\omega}_{ci}}}

\newcommand{\kratraw}{\kpar/\kperp}
\newcommand{\krat}{\abs{\kratraw}}
\newcommand{\rhob}{\rho_{\perp b}}

\newcommand{\zp}{\zeta}

\newcommand{\eqlab}[1]{\quad\text{#1}}
\newcommand{\CAElab}{\eqlab{CAE}}
\newcommand{\GAElab}{\eqlab{GAE}}

\newcommand{\ftail}{f_\text{tail}}

\newcommand{\flr}{\xi}

\newcommand{\alphamin}{\alpha_\text{min}}
\newcommand{\alphamax}{\alpha_\text{max}}

\newcommand{\avg}[1]{\left\langle #1 \right\rangle}
\renewcommand{\vec}[1]{\bm{#1}}

\newcommand{\abs}[1]{\left|#1\right|}
\renewcommand{\dot}{\cdot}

\newcommand{\defined}{\equiv}
\newcommand{\like}{\sim}

\newcommand{\ord}[1]{\mathcal{O}\left(#1\right)}
\newcommand{\plusord}[1]{\, + \, \ord{#1}}

\newcommand{\approptoinn}[2]{\mathrel{\vcenter{
  \offinterlineskip\halign{\hfil$##$\cr
    #1\propto\cr\noalign{\kern2pt}#1\sim\cr\noalign{\kern-2pt}}}}}
\newcommand{\appropto}{\mathpalette\approptoinn\relax}

\newcommand{\tto}{\text{ to }}

\newcommand\numberthis{\addtocounter{equation}{1}\tag{\theequation}}

\newcommand{\figref}[1]{Fig.\xspace\ref{#1}}
\renewcommand{\eqref}[1]{Eq.\xspace\ref{#1}}
\newcommand{\secref}[1]{Sec.\xspace\ref{#1}}
\newcommand{\citeref}[1]{Ref.\xspace\onlinecite{#1}}
\newcommand{\appref}[1]{Appendix\xspace\ref{#1}}
\newcommand{\tabref}[1]{Table\xspace\ref{#1}}

\newcommand{\code}[1]{\texttt{#1}\xspace}
\newcommand{\HYM}{\code{HYM}}
\newcommand{\TRANSP}{\code{TRANSP}}
\newcommand{\NUBEAM}{\code{NUBEAM}}

\newcommand{\myname}{J.B. Lestz} 
\newcommand{\Elena}{E.V. Belova} 
\newcommand{\Nikolai}{N.N. Gorelenkov} 
\newcommand{\Neal}{N.A. Crocker} 
\newcommand{\Shawn}{S.X. Tang}

\newcommand{\PPPL}{Princeton Plasma Physics Lab, Princeton, NJ 08543, USA}
\newcommand{\Princeton}{Department of Astrophysical Sciences, Princeton University, Princeton, NJ 08543, USA}
\newcommand{\UCLA}{Department of Physics and Astronomy, University of California, Los Angeles, CA 90095, USA}

\newcommand{\Podesta}{Podest{\`a}}
\newcommand{\Fulop}{F{\"u}l{\"o}p}

\definecolor{darkgreen}{rgb}{0,0.5,0}

\begin{document}

\title{Analytic stability boundaries for compressional and global \Alfven eigenmodes driven by fast ions. II. Interaction via Landau resonance.}
\author{\myname}
\email{jlestz@pppl.gov}
\affiliation{\Princeton}
\affiliation{\PPPL}
\author{\Nikolai}
\affiliation{\PPPL}
\author{\Elena} 
\affiliation{\PPPL}
\author{\Shawn}
\affiliation{\UCLA}
\author{\Neal}
\affiliation{\UCLA}
\date{\today}
\begin{abstract}
Conditions for net fast ion drive are derived for beam-driven, co-propagating, sub-cyclotron compressional (CAE) and global (GAE) \Alfven eigenmodes driven by the Landau resonance with super-\Alfvenic fast ions. Approximations applicable to realistic neutral beam distributions and mode characteristics observed in spherical tokamaks enable the derivation of marginal stability conditions for these modes. Such conditions successfully reproduce the stability boundaries found from numerical integration of the exact expression for local fast ion drive/damping. Coupling between the CAE and GAE branches of the dispersion due to finite $\omeganorm$ and $\krat$ is retained and found to be responsible for the existence of the GAE instability via this resonance. Encouraging agreement is demonstrated between the approximate stability criterion, simulation results, and a database of NSTX observations of co-CAEs. 
\end{abstract}
\maketitle
\section{Introduction}
\label{sec:introduction}

High frequency co-propagating compressional \Alfven eigenmodes (CAE) have been observed in the spherical tokamaks NSTX(-U)\cite{Fredrickson2001PRL,Fredrickson2013POP,Fredrickson2019POP} and MAST.\cite{Appel2008PPCF,Sharapov2014PP,McClements2017PPCF} These instabilities are more easily excited on these devices than conventional tokamaks due to their lower magnetic fields and large neutral beam power, which together generate a substantial population of super-\Alfvenic fast ions ($\vinj  = 2 - 6$).\cite{McClements2017PPCF}. They are typically observed with frequencies of $\omeganorm = 0.3 - 1.2$ and toroidal mode numbers $\abs{n} = 3 - 15$. CAEs are the compressional MHD wave, with approximate dispersion $\omega = k\va$ in a uniform, zero beta slab, where $\va = B/\sqrt{\mu_0 n_i  m_i}$ is the \Alfven speed. They are polarized with finite $\vdB\dot\vkperp$ and $\dbpar$. In tokamak geometry, the mode becomes confined within an effective potential well\cite{Mahajan1983bPF,Gorelenkova1998POP,Kolesnichenko1998NF,Gorelenkov2002POP,Gorelenkov2002NF,Smith2003POP,Gorelenkov2006NF,Smith2009PPCF,Smith2017PPCF} with discrete frequencies resulting from the boundary conditions. 

GAEs are a class of weakly damped shear MHD waves that can exist just below or above\cite{Kolesnichenko2007POP} an extremum in the continuum of solutions for shear \Alfven waves satisfying $\omega = \abs{\kpar(r)}\va(r)$. In contrast, modes within the \Alfven continuum are rapidly sheared apart by phase mixing, and therefore are rarely observed in experiments.\cite{Heidbrink2008POP,Gorelenkov2014NF} Co-propagating GAEs were initially modeled numerically in cylindrical plasmas\cite{Ross1982PF,Appert1982PP} in order to explain resonant peaks in antenna loading in the TCA tokamak.\cite{DeChambrier1982POP} Further theoretical work found them to be stabilized by finite toroidicity effects\cite{Fu1989PF,VanDam1990FT} in the limit of $\omeganorm \ll 1$. The shear waves are polarized such that $\vdB \dot\vk = 0$ and $\dbpar = 0$ in a uniform plasma. The discrete spectrum of GAEs exists due to coupling to the magnetosonic mode, an equilibrium current, current density gradient, and finite $\omeganorm$ effects.\cite{Appert1982PP,Mahajan1983PF,Mahajan1984PF,Li1987PF,Fu1989PF,VanDam1990FT} Excitation of CAEs/GAEs requires a resonant population of energetic particles with sufficient velocity space gradients to overcome background damping sources. In this work, the Landau (non-cyclotron) resonance is considered. Interaction via the ordinary and anomalous cyclotron resonances were studied in part 1.\cite{Lestz2019p1}

The stability of NBI-driven CAEs due to the Landau resonance has previously been studied by Belikov\cite{Belikov2003POP,Belikov2004POP} in application to NSTX. In those works, a delta function distribution in pitch was assumed for the fast ions, which is an unrealistic model for the typically broad distributions inferred from experimental modeling. Previous works also assumed $\kpar \ll \kperp$ and $\omega \ll \omegaci$, whereas experimental observations and simulations demonstrate that $\kpar \like \kperp$ and $\omega \like \omegaci/2$ are common. Here, prior work is extended to provide a local expression for the fast ion drive due to a general beam-like distribution through the Landau resonance. Terms to all order in $\omeganorm$ and $\krat$ are kept for applicability to the entire possible spectrum of modes. In particular, finite $\omeganorm$ and $\krat$ introduces coupling between the compressional and shear branches of the dispersion which enables the GAE to be excited through this resonance. Full finite Larmor radius (FLR) terms are also retained, similar to prior studies. As in part 1 for the cyclotron resonances,\cite{Lestz2019p1} experimentally relevant regimes have been identified where approximate stability boundaries can be derived. Since other damping sources have not been included in this work, the derived conditions for net fast ion should be treated as necessary but insufficient conditions for instability. 

The paper is structured as follows. The fast ion drive for CAEs/GAEs from the Landau resonance is derived analytically in the local approximation in \secref{sec:derivation}, based on the framework in \citeref{Mikhailovskiiv6} and applied to a parametrized neutral beam distribution. Approximations are made to this expression in \secref{sec:approxstab} in order to derive marginal stability conditions in the limits of very narrow (\secref{sec:narrow}) and realistically broad (\secref{sec:wide}) fast ion distributions. Within \secref{sec:wide}, the limits of small and large FLR effects are treated separately in \secref{sec:slow} and \secref{sec:fast}, respectively, and the dependence of the drive/damping on fast ion parameters for fixed mode properties is discussed and compared to the approximate analytic conditions. A complementary discussion of the fast ion drive/damping as a function of the mode properties for fixed fast ion parameters is presented in \secref{sec:stabao}, including the role of compressional-shear mode coupling in setting the stability boundaries. A comparison of the approximate stability conditions against a database of co-CAE activity in NSTX is shown in \secref{sec:expcomp}. Lastly, a summary of the main results and discussion of their significance is given in \secref{sec:summary}. 

\section{Fast Ion Drive for General Beam Distribution in the Local Approximation for the Landau Resonance} 
\label{sec:derivation}

As in part 1,\cite{Lestz2019p1} we note that due to the large frequencies of these modes in experiments: $\omeganorm = 0.3 \tto 1$ and $\krat$ often order unity in simulations, it is worthwhile to consider the dispersion relation for the shear and compressional branches including coupling due to thermal plasma two fluid effects. Additional coupling can be induced by spatial gradients present in realistic experimental profiles, which is not included in our analysis. 

\subsection{Derivation}

Define $\omegabar = \omega/\omegacio$, $N = k\va/\omega$, $A = (1 - \omegabar^2)^{-1}$, and also $F^2 = \kpar^2/k^2$, $G = 1 + F^2$. Here, $\omegacio$ is the on-axis ion cyclotron frequency. Then in uniform geometry, the local dispersion in the MHD limits of $E_\parallel \ll E_\perp$ and $\omega \ll \abs{\omegace},\omegape$ is\cite{Stix1975NF} 

\begin{equation}
N^2 = \frac{AG}{2F^2}\left[1 \pm \sqrt{1 - \frac{4F^2}{AG^2}}\right]
\label{eq:stixdisp}
\end{equation}

The ``$-$" solution corresponds to the compressional \Alfven wave (CAW), while the ``$+$" solution corresponds to the shear \Alfven wave (SAW). The coupled dispersion can modify the polarization of the two modes relative to the uncoupled approximation. In \citeref{Lestz2019p1}, it was shown that the growth rates for the cyclotron resonance-driven cntr-CAEs and co-GAEs have local maxima with respect to $\krat$, whereas they increase monotonically when this coupling is neglected. The low frequency approximation of \eqref{eq:stixdisp} is $\omega \approx k\va$ for CAEs and $\omega \approx \abs{\kpar}\va$ for GAEs, which can simplify analytic results when valid. The Landau resonance describes a wave-particle interaction satisfying the following relation

\begin{align}
\omega - \avg{\kpar\vpar} - \avg{\kperp\vdrift} = 0
\label{eq:rescon}
\end{align}

Above, $\avg{\dots}$ denotes poloidal orbit averaging appropriate for the ``global'' resonance (see further discussion in part 1\cite{Lestz2019p1} and also \citeref{Belikov2003POP}). As in part 1, we consider the approximation of $\abs{\kperp\vdrift} \ll \abs{\kpar\vpar}$, focusing on the primary resonance and neglecting sidebands. Hence, all modes satisfying this resonance with particles with $\vpres \defined \avg{\vpar} > 0$ must be co-propagating with $\kpar > 0$. 

The stability calculation will be applied to the same model fast ion distribution as in part 1, motivated by theory and \NUBEAM modeling of NSTX discharges,\cite{Belova2017POP} written as a function of $v = \sqrt{2\W/m_i}$ and $\lambda = \mu B_0 / \W$ in separable form: $\fb(v,\lambda)= C_f n_b f_1(v)f_2(\lambda)$, defined below

\begin{subequations}
\begin{align}
\label{eq:F1}
f_1(v) &= \frac{\ftail(v;v_0)}{v^3 + v_c^3} \\ % \eqlab{for $v < v_0$} \\
\label{eq:F2}
f_2(\lambda) &= \exp\left(-\left(\lambda - \lambda_0\right)^2 / \Delta\lambda^2\right)
\end{align}
\label{eq:Fdistr}
\end{subequations}

The constant $C_f$ is for normalization. The first component $f_1(v)$ is a slowing down function in energy with a cutoff at the injection energy $v_0$ and a critical velocity $v_c$, with $\ftail(v;v_0)$ a step function. The second component $f_2(\lambda)$ is a Gaussian distribution in $\lambda$. To lowest order in $\mu \approx \mu_0$, it can be re-written as $\lambda = (\vperp^2/v^2)(\omegacio/\omegaci)$. The distribution in the final velocity component, $\pphi$, is neglected in this study for simplicity, as it is expected to be less relevant for the high frequencies of interest for these modes. NSTX typically operated with $\vinj = 2 - 6$ and $\linj = 0.5 - 0.7$ with $\dl = 0.3$. Early operations of NSTX-U had $\vinj < 3$, featuring an additional beam line with $\linj \approx 0$. For this study, $v_c = v_0/2$ is used as a characteristic value. 

In part 1, the fast ion drive/damping was derived perturbatively in the local approximation for a two component plasma comprised of a cold bulk plasma and a minority hot ion population. Restricting Eq. 21 of \citeref{Lestz2019p1} to the $\lres = 0$ Landau resonance and applying to the model distribution gives 

\begin{widetext}
\begin{multline}
\frac{\gamma}{\omegaci} = -\frac{n_b}{n_e}\frac{\pi C_f v_0^3 \bres^{3/2}}{2 v_c^3\omegabar} 
\left\{ \int_0^{1-\bres} \frac{x \Jzm(\flr(x,\zp))}{(1-x)^2}\frac{e^{-(x-\xinj)^2/\dx^2}}{1 + \frac{v_0^3}{v_c^3}\left(\frac{\bres}{1-x}\right)^{3/2}}
\left[-\frac{x(x-\xinj)}{\dx^2} + \frac{3}{4}\frac{1}{1 + \frac{v_c^3}{v_0^3}\left(\frac{1-x}{\bres}\right)^{3/2}}\right]dx \right. \\  
\left.\vphantom{\left[\frac{3}{4}\frac{1}{1 + \left(\frac{1-x}{4\bres}\right)^{3/2}}\right]}
+ \frac{\bres^{-1}-1}{2\left(1 + \frac{v_0^3}{v_c^3}\right)}e^{-(1 - \bres-\xinj)^2/\dx^2}\Jzm\left(\zp\sqrt{\bres^{-1}-1}\right)\right\}
\label{eq:gammabeam}
\end{multline}
\end{widetext}

All notation is the same as defined in part 1. Briefly, the integration variable is $x = \vperp^2/v^2 = \lambda\omegacires$. Similarly, $\xinj = \linj\omegacires$ and $\dx = \dl\omegacires$. The resonant parallel energy fraction is $\bres = \vpres^2/v_0^2$. \eqref{eq:gammabeam} is valid for arbitrary $\omeganorm$ and $\krat$, generalizing results published in \citeref{Belikov2003POP,Belikov2004POP} for the co-CAE driven by the Landau resonance, which were restricted to $\omega \ll \omegaci$ and $\kpar \ll \kperp$. This generalization is contained mostly in the FLR effects, within the function $\Jzm(\flr)$, defined for arbitrary $\lres$ in Eq. 16 of part 1, which simplifies for $\lres = 0$ to 

\begin{align}
\label{eq:Jlm}
\Jzm(\flr) &= \frac{N^{-2}\left(N^{-2} - F^2(1 - \omegabar^2)\right)}{N^{-4} - F^2}J_1^2(\flr) \\ 
\text{where } \label{eq:zsimp}
\flr &= \kperp\rhob \defined \zp \sqrt{\frac{x}{1-x}} \\ 
\text{and }
\zp &= \frac{\kperp\vpres}{\omegaci} = \frac{\omegabar}{\alpha}
\label{eq:zp}
\end{align}

Above, $\rhob = \vperp/\omegaci$ is the Larmor radius of the fast ions, $\flr$ is the FLR parameter, and $\zp$ is the modulation parameter describing how rapidly the integrand of \eqref{eq:gammabeam} oscillates, which depends on the mode parameters $\omegabar = \omeganorm$ and $\alpha \defined \krat$. The $m$ in $\Jzm(\flr)$ denotes the mode dispersion (= `$C$' for CAE and `$G$' for GAE), as contained within $N$ (given in \eqref{eq:stixdisp} for CAEs using the ``-" solution and GAEs using the ``+'' solution). As argued in part 1,\cite{Lestz2019p1} $\Jzm(\flr) \geq 0$ for both modes. In the limit of $\omeganorm \ll 1$, 

\begin{subequations}
\label{eq:Jlmsmallappx}
\begin{align}
\label{eq:Jlmsmallappx-cae}
\lim_{\omegabar\rightarrow 0} \Jzc(\flr) &= J_1^2(\flr)
\CAElab \\ 
\label{eq:Jlmsmallappx-gae}
\lim_{\omegabar\rightarrow 0} \Jzg(\flr) &= 
\omegabar^2\alpha^4 J_1^2(\flr)
\GAElab
\end{align}
\end{subequations}

Hence, GAEs may only interact with fast ions via the Landau resonance when finite $\omeganorm$ and $\krat$ are considered. In another limit, where $0 < \omegabar < 1$ and $\alpha \gg 1$, FLR function reduces to 

\begin{align}
\lim_{\alpha\rightarrow \infty} \Jzm(\flr) &= \frac{\left(1 \pm \omegabar\right)^2}{2\pm \omegabar}J_1^2(\flr) 
\label{eq:Jlmbigappx}
\end{align}

In \eqref{eq:Jlmbigappx}, the top signs are for CAEs, and the bottom signs for GAEs. The expression in \eqref{eq:gammabeam} represents the local perturbative growth rate for CAE/GAEs in application to a general beam-like distribution of fast ions, keeping all terms from $\omeganorm$, $\krat$, and $\kperp\rhob$. The derivative with respect to $\pphi$ has been omitted, as it is expected to less relevant for the high frequency modes studied here. Moreover, the local approximation ignores spatial profile dependencies, sacrificing accuracy in the magnitude of the growth/damping rate in favor of deriving more transparent instability conditions. 

\subsection{Properties of Fast Ion Drive}

Notice that only regions of the integrand where the term in brackets is negative are driving. For modes interacting via the Landau resonance, this requires $\partial\fb/\partial\lambda < 0$, equivalent to $\lambda > \linj$ for a distribution peaked at $\linj$. Unlike the cyclotron resonance-driven modes analyzed in part 1, the damping from $\partial \fb/\partial v$ (second term in square brackets) can be comparable to the drive/damping from velocity space anisotropy over a nontrivial fraction of the integration region. Consequently, an immediate stability condition can be extracted. 

When $1 - \vpres^2/v_0^2 \leq \linj\omegacires$, the integrand is non-negative over the region of integration, such that $\gamma < 0$. As a corollary, when $1 - \vpres^2/v_0^2 \leq \linj\omegacires$, modes interacting through the Landau resonance are strictly stable. For CAEs, $\vpres$ depends on $\krat$, and hence this relation provides information about the allowed mode properties driven by a given distribution of fast ions. 

As mentioned above, the co-GAE instability due to the Landau resonance is possible only when coupling to the compressional branch is considered. Neglecting coupling, its FLR function $\J{0}{G}$ would be identically zero according to \eqref{eq:Jlm} in the limit of $N^{-2} = F^2 (1 - \omegabar^2)$ exactly. However, even when considering the coupling, its growth rate is much smaller compared to the co-CAE due to the additional factor of $\omegabar^2\alpha^4$ in \eqref{eq:Jlmsmallappx-gae}, which is typically small for $\omegabar < 1$ and $\alpha \like 1$. Consequently, the co-GAE will be at most weakly unstable due to this resonance, and perhaps stabilized entirely by electron Landau or continuum damping.\cite{Fu1989PF} In contrast, co-CAEs have less barriers to excitation, consistent with their measurement in NSTX\cite{Fredrickson2001PRL,Fredrickson2013POP} and MAST,\cite{Appel2008PPCF,Sharapov2014PP} and also their appearance in \HYM modeling of NSTX.\cite{Belova2017POP} Both instabilities require finite $\kperp\rhob$ for excitation, since their FLR functions are $\J{0}{m} \propto J_1^2(\flr) \rightarrow 0$ for $\flr \rightarrow 0$. 

The expression for growth rate in \eqref{eq:gammabeam} also demonstrates that instability can occur for any value of $\kperp\rhob > 0$, depending on the parameters of the fast ion distribution. This extends the region of instability found for co-CAEs driven by passing particles in \citeref{Belikov2003POP}, which asserted that $\sqrt{\linj}(\omeganorm)(\vinj) < 2$ was necessary for instability, due to the additional assumption of a delta function distribution in $\lambda$. Similarly, the conclusions from the same authors in \citeref{Belikov2004POP} regarding co-CAE stabilization by trapped particles, while qualitatively consistent with the findings here, are likewise limited to the case of a vanishingly narrow distribution in $\lambda$. For further understanding of the relationships between the relevant parameters required for instability, analytic approximations or numerical methods must be employed. 

\begin{figure*}[tb]
\hspace*{-2ex}
\subfloat[]{\includegraphics[width = 0.365\textwidth]{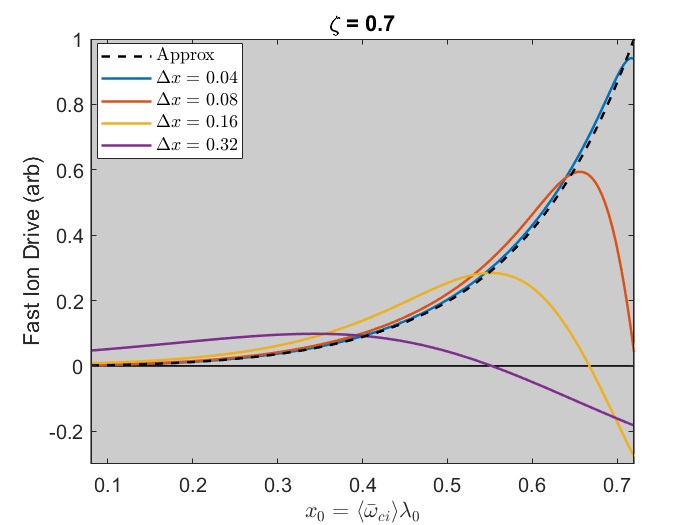}}
\hspace*{-4ex}
\subfloat[]{\includegraphics[width = 0.365\textwidth]{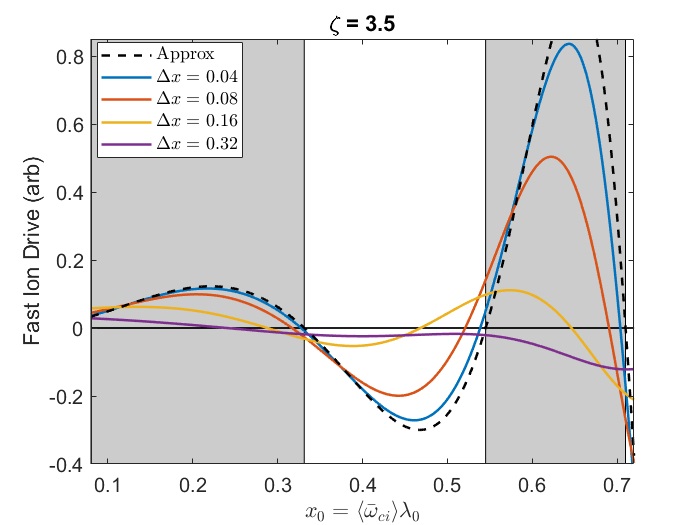}}
\hspace*{-4ex}
\subfloat[]{\includegraphics[width = 0.365\textwidth]{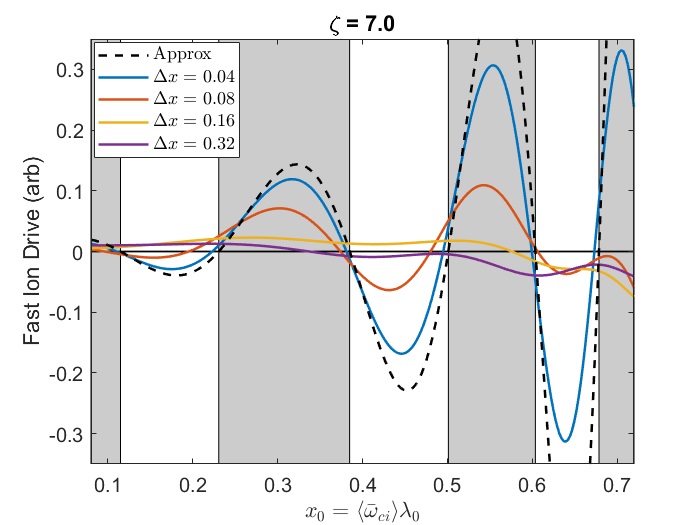}}
\caption{Comparison of numerically integrated growth rate to narrow beam approximation for co-CAEs/GAEs with $\bres = 0.2$ as a function of the central trapping parameter of the beam distribution. Black dashed line shows the analytic approximation made in \eqref{eq:gammasimp} for $\dx = 0.04$ and (a) $\zp = 0.7$, (b) $\zp = 3.5$, and (c) $\zp = 7.0$. Colored curves show numerical integration of \eqref{eq:gammabeam} for different values of $\dx$: blue $\dx = 0.04$, orange $\dx = 0.08$, gold $\dx = 0.16$, and purple $\dx = 0.32$. Shaded regions correspond to regions of drive according to the narrow beam approximation.}
\label{fig:narrowcomp}
\end{figure*}

\section{Approximate Stability Criteria} 
\label{sec:approxstab}

The expression derived in \eqref{eq:gammabeam} can not be integrated analytically, and has complicated parametric dependencies on properties of the specific mode of interest: GAE vs CAE, $\krat$, $\omeganorm$, as well as on properties of the fast ion distribution: $\vinj$, $\linj$, and $\dl$. For chosen values of these parameters, the net fast ion drive can be rapidly calculated via numerical integration. Whenever $1 - \vpres^2/v_0^2 \leq \linj\omegacires$, both modes are damped via the Landau resonance provided that the fast ion distribution is monotonically decreasing in energy (\eg slowing down) and has a single peak in $\lambda$ at $\lambda = \linj$, such as the beam distribution given in \eqref{eq:Fdistr}. There are also regimes where approximations can be made in order to gain insight into the stability properties analytically: one where the fast ion distribution is very narrow ($\dl \lesssim 0.10$) and one where it is moderately large $(\dl \gtrsim 0.20$). The former allows comparison with previous calculations,\cite{Belikov2003POP,Belikov2004POP} while the latter includes the experimental regime where the distribution width in NSTX is typically $\dl \approx 0.30$. In this section, marginal stability criteria will be derived in these regimes and compared to the numerical evaluation of \eqref{eq:gammabeam}, using $v_c = v_0/2$ and $n_b/n_e = 5.3\%$, based on the conditions in the well-studied NSTX H-mode discharge $\# 141398$. 

\subsection{Approximation of Very Narrow Beam}
\label{sec:narrow}

For the first regime, consider the approximation of a very narrow beam in velocity space. The purpose of this section is to determine when such an approximation can correctly capture the sign of the growth rate. Hence, consider $\dx \ll 1$ such that only a small region $\xinj - \delta < x < \xinj + \delta$ contributes to the integral, where $\delta \approx 2\dx$. So long as $0 < \xinj - \delta$ and $\xinj + \delta < 1 - \bres$, two linear approximations can be made such that to leading order in $\dx$, \eqref{eq:gammabeam} is approximately 

\begin{align}
\label{eq:gammasimp}
\frac{\gamma}{\omegaci} &\propto C_f \dx\sqrt{\pi}\left[2 h_1'(\xinj) - 3 h_2(\xinj)\right] \\
\text{where } h_1(x) &= \frac{x^2}{(1-x)^2}\frac{\Jzm(\flr(x,\zp))}{1 + \frac{v_0^3}{v_c^3}\left(\frac{\bres}{1-x}\right)^{3/2}} \\ 
%\text{and } h_2(x) &= \frac{x}{(1-x)^2}\frac{\Jzm(\flr(x,\zp))}{1 + \frac{v_0^3}{v_c^3}\left(\frac{\bres}{1-x}\right)^{3/2}}
%\frac{1}{1 + \frac{v_c^3}{v_0^3}\left(\frac{1-x}{\bres}\right)^{3/2}}
\text{and } h_2(x) &= \frac{h_1(x)}{x}\frac{1}{1 + \frac{v_c^3}{v_0^3}\left(\frac{1-x}{\bres}\right)^{3/2}}
\end{align} 

The above expressions apply equally to CAEs and GAEs. Whereas for the cyclotron resonances discussed in part 1, the narrow beam approximation yielded a growth rate with sign depending only on the sign of a single function,\cite{Lestz2019p1} for the Landau resonance, a second function must be kept to include the non-negligible contribution from $\partial\fb/\partial v$. A comparison of the approximate narrow beam stability criteria to the exact expression with $\bres = 0.2$ is shown in \figref{fig:narrowcomp}. There, the dashed line shows the approximate analytic result \eqref{eq:gammasimp} plotted as a function of $\xinj$ for $\dx = 0.04$ and different values of $\zp$. Values of $\xinj$ where $\gamma > 0$ according to \eqref{eq:gammasimp} indicate regions where the fast ions are net driving according to this assumption (shaded regions). For comparison, the full expression \eqref{eq:gammabeam} is integrated numerically for each value of $\xinj$ for varying $\dx = 0.04, 0.08, 0.16, 0.32$. 
% While the magnitudes are not expected to match since the magnitude of the analytic result also depends on $\dx$, 
This figure demonstrates where the narrow beam approximation correctly determines the sign of the fast ion drive, and how it depends on $\zp$. As in part 1 for cntr-GAEs driven by the ordinary cyclotron resonance, it is demonstrated that $\dx \approx 0.1$ gives an acceptable (albeit strained) agreement between the approximation and numerically integrated expression. For any larger values (such as $\dx = 0.16$ and $\dx = 0.32$ shown), the approximation no longer captures the correct sign of the growth rate as a function of $\xinj$, with more pronounced disagreement occurring at larger values of $\xinj$. Moreover, it is clear that larger $\zp$ leads to more distinct regions of net drive and damping, leading to more areas where the approximate formula may incorrectly predict stability or instability. 

\subsection{Approximation of Realistically Wide Beam}
\label{sec:wide}

For sufficiently wide beam distributions (such as those generated with NBI in NSTX with $\dx \approx 0.3$), one may approximate $d \exp(-(x-\xinj)^2/\dx^2)/dx \approx -2(x-\xinj)/\dx^2$. This linear approximation is appropriate for $\xinj - \dx/\sqrt{2} < x < \xinj + \dx/\sqrt{2}$. When this range extends over a large fraction of the integration region, it can be used to provide very accurate marginal stability conditions. Throughout this section, $v_c = v_0/2$ will be taken as a representative figure, and the slowing down part of the distribution will be approximated as constant since it makes a small quantitative difference. However, this approximation alone is insufficient to evaluate \eqref{eq:gammabeam} in terms of elementary functions, as the Bessel functions with complicated arguments remain intractable. 

For the cyclotron resonances analyzed in part 1, the fast ion damping due to $\partial\fb/\partial v$ could be neglected since it was smaller than the drive/damping due to $\partial\fb/\partial \lambda$ in that case by a factor of $\omegabar\dx^2 \ll 1$ except in a very small region near $x = \xinj$. For modes driven by the Landau resonance, it can compete with the drive/damping from anisotropy over a wider range of the integration region. Hence, the contributions from $\partial\fb/\partial v$ must be kept in this section, leading to somewhat more complicated instability boundaries than those derived in Sec IV B of part 1.

Substituting the values of $\omeganorm$ and $\krat$ from the most unstable modes in \HYM simulations into \eqref{eq:zp} shows that the majority of these modes have $\zp \like \ord{1}$. Since this parameter controls how rapidly $\Jlm(\flr)$ oscillates, we are motivated to consider two cases separately: $\zp \ll 1$ (small FLR, more common) and $\zp \gg 1$ (large FLR, uncommon for $\omega < \omegaci$). 

\subsubsection{Small FLR regime \texorpdfstring{$(\zp \ll 1)$}{}}
\label{sec:slow}

Consider first the case of small FLR effects. For small argument, $\Jlm(\flr) \propto J_1^2(\flr) \approx \flr^2/4 \plusord{\flr^4}$. Then the simplified integral to consider is 

\begin{multline}
\gamma \appropto \int_0^{1-\bres} \frac{x^3(x-\xinj)}{(1-x)^3}dx \\ 
- \frac{3\dx^2}{4}\int_0^{1-\bres} \frac{x^2}{(1-x)^3}\frac{dx}{1 + \left(\frac{1-x}{4\bres}\right)^{3/2}} \\ 
- \frac{\dx^2}{2}\left(\bres^{-1}-1\right)^2 e^{-(1-\bres-\xinj)^2/\dx^2}
\label{eq:CAEdampgam}
\end{multline}

As a reminder, $\eta = \vpres^2/v_0^2$ such that the upper bound of the integration describes a cutoff in the distribution function at the finite injection velocity $v_0$. The integrals can be evaluated exactly and well-approximated. Solving for the marginal stability condition $\gamma = 0$, neglecting the third term for now, yields 

\begin{align}
\label{eq:CAEdampgamsym}
\xinj &= g_0(\bres) -\dx^2 g_1(\bres) \\ 
\label{eq:CAExcrit}
 &\approx 1 - \bres^{4/5} - \frac{2\dx^2}{3(1 - \bres^{4/5})} \\
\label{eq:wideslowCAEgam}
\Rightarrow v_0 &= \frac{\vpres}{\left[1 - \frac{1}{2}\left(\xinj + \sqrt{\xinj^2 + \frac{8\dx^2}{3}}\right)\right]^{5/8}}
\end{align}

The exact forms of $g_0(\bres)$ and $g_1(\bres)$ are given in \appref{app:dampref}. The first function can be excellently approximated by $g_0(\bres) \approx 1 - \bres^{4/5}$, with a maximum relative error of less than $1\%$. The second function, $g_1(\bres)$ is substantially more complicated. Noting its singularity as $\bres\rightarrow 1$, and considering that the goal is to find a closed form for $\bres$ as a function of $\xinj$, an \emph{ansatz} of the form $c/(1 - \bres^{4/5})$ is chosen, with $c = 2/3$ giving a maximum relative error of $15\%$, and usually half that. With this approximation, the marginal stability condition could be derived. 

When $\dx$ is small, \eqref{eq:wideslowCAEgam} would reduce to $v_0 = \vpres/(1 - \xinj)^{5/8}$, similar in form to the marginal stability condition found in part 1 for cyclotron resonance-driven modes with $\zp \ll 1$, except with a power of $5/8$ instead of $3/4$ due to the different $\Jlm(\flr)$ functions. Note that $\gamma < 0$ for \emph{all} values of $\xinj,v_0$ when $\dx^2 > 5/3$ according to \eqref{eq:CAExcrit}. This condition represents the beam width necessary to balance the maximum anisotropy drive with the slowing down damping. While it indicates a theoretical avenue for stabilizing all CAEs/GAEs driven by the Landau resonance, it is unlikely to be useful in practice since it requires a nearly uniform distribution in $\lambda$, which would not allow sufficient flexibility in the current profile that is desirable for other plasma performance objectives. 

The third term in \eqref{eq:CAEdampgam} was neglected because its inclusion would prevent an algebraic solution for $\xinj$ at marginal stability. However, it can be comparable in magnitude to the second term in the integration, and can be included in an \emph{ad-hoc} fashion by solving for its effect at $\xinj = 0$, and multiplying the full result by this factor. We will also apply a rational function approximation to the Gaussian dependence, so that at $\xinj = 0$, the marginal stability condition for $\bres$ is 

\begin{multline}
\label{eq:bdampcrit}
\int_0^{1-\bres} \frac{x^4}{(1-x)^3}dx \\ 
- \frac{3\dx^2}{4}\int_0^{1-\bres} \frac{x^2}{(1-x)^3}\frac{dx}{1 + \left(\frac{1-x}{4\bres}\right)^{3/2}} \\ 
- \frac{\dx^4}{2}\frac{\left(\bres^{-1}-1\right)^2}{\dx^2 + (1-\bres)^2} = 0
\end{multline}

This expression yields a quadratic formula for $\dx^2$, given in \appref{app:dampref}, which can be approximated to within $10\%$ globally and inverted to yield 

\begin{align}
\bres \approx (1 - \dx^{4/5})^{5/4}
\end{align}

Hence, the modification to the marginal stability condition necessary to match the correction due to the third term in \eqref{eq:CAEdampgam} at $\xinj = 0$ is 

\begin{align}
v_0 &= \frac{\vpres}{\left[1 - \frac{1}{2}\left(\xinj + \sqrt{\xinj^2 + \frac{8\dx^2}{3}}\right)\right]^{5/8}}\left(\frac{1 - \dx\sqrt{2/3}}{1-\dx^{4/5}}\right)^{5/8}
\label{eq:wideslowCAEgamtail}
\end{align}

\begin{figure}[tb]
\subfloat[\label{fig:wideslowCAE}]{\includegraphics[width = \halfwidth]{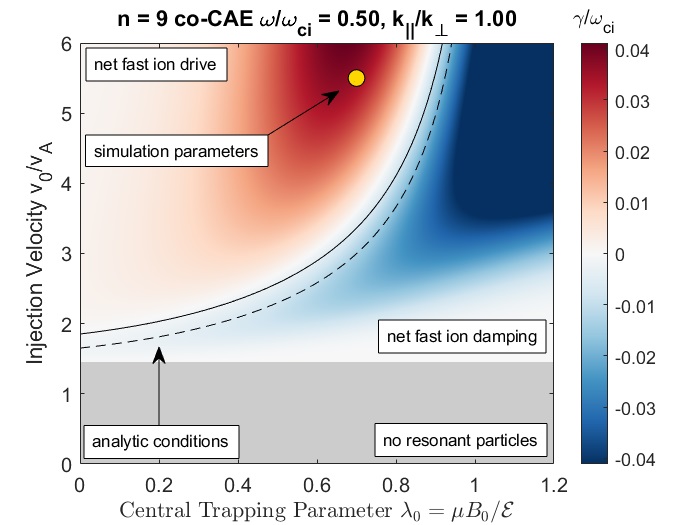}} \\
\subfloat[\label{fig:wideslowGAE}]{\includegraphics[width = \halfwidth]{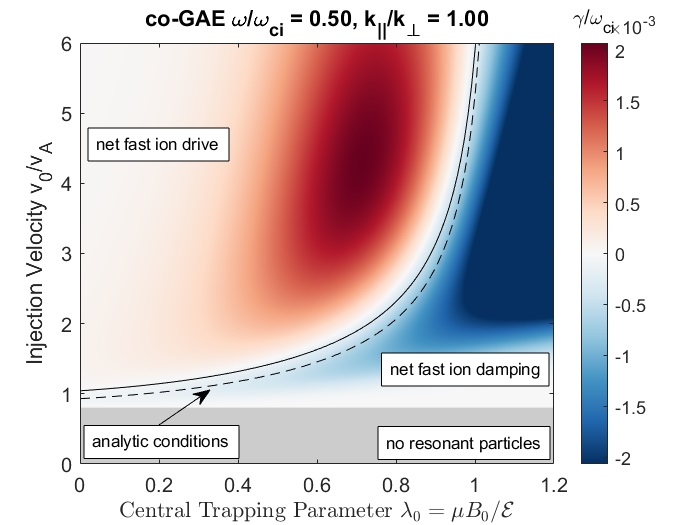}} \\
\caption{Numerical integration of full growth rate expression \eqref{eq:gammabeam} as a function of fast ion distribution parameters $\vinj$ and $\linj$ with $\dx = 0.30$ for a Landau-resonance driven (a) co-CAE and (b) co-GAE in the small FLR regime ($\zp \ll 1$) with properties inferred from \HYM simulations: $\omeganorm = 0.5$ and $\krat = 1$, implying $\zp = 0.5$. Red indicates net fast ion drive, blue indicates net fast ion damping, and gray indicates beam parameters with insufficient energy to satisfy the resonance condition. Dashed curve shows approximate stability condition excluding damping from the tail, derived in \eqref{eq:wideslowCAEgam}. Solid curve shows approximate stability condition including damping from the tail, derived in \eqref{eq:wideslowCAEgamtail}.}
\label{fig:wideslowfig}
\end{figure}

This marginal stability condition encompasses both the CAEs and GAEs, since the only difference is that the GAEs have a reduced magnitude, as described by \eqref{eq:Jlmsmallappx} and \eqref{eq:Jlmbigappx} when $\omegabar \ll 1$ and $\alpha \gg 1$, respectively. The condition derived in \eqref{eq:wideslowCAEgamtail} can also be compared against the full numerically integrated expression in 2D beam parameter space for a typical case, shown in \figref{fig:wideslowCAE}. There, an $n = 9$ co-CAE driven by the Landau resonance in \HYM simulations has been chosen, using mode parameters of $\omeganorm = 0.5$ and $\krat = 1$, implying $\zp = 0.5$ and a distribution with $\dx = 0.30$. There, the solid curve includes the contribution from the tail of the distribution (\eqref{eq:wideslowCAEgamtail}), while the dashed curve neglects this contribution (\eqref{eq:wideslowCAEgam}). The former better tracks the numerically computed stability boundary. Note also that the boundary is shifted upwards due to the damping from including the velocity derivative terms.  

It is worth pointing out that the Landau resonance co-CAEs require relatively large $\vinj$ for excitation compared to the cntr-propagating modes driven by the ordinary cyclotron resonance. To see this, consider \eqref{eq:wideslowCAEgamtail} and substitute $\vpres \approx k/\abs{\kpar}$, which follows from the approximate dispersion $\omega \approx k\va$ for CAEs. Then the minimum $\vinj$ for instability occurs at $\xinj = 0$, such that $\vinj > \abs{k/\kpar}/(1 - \dx^{4/5})^{5/8}$. This expression in turn is minimized for $\abs{k/\kpar} \rightarrow 1$, which for $\dx = 0.3$ yields $\vinj > 1.3$ as a strict lower bound for this instability. With more realistic perpendicular beam injection, such as the original NSTX beam with $\linj \approx 0.7$, the requirement increases to $\vinj > 2.9$ in the same limit of $\abs{k/\kpar} \rightarrow 1$, and even larger at $\vinj > 4.1$ for common values of $\kpar/\kperp \approx 1$. 

In contrast, cyclotron resonance-driven cntr-GAE excitation features no such constraints, as modes can in principle be excited even for $\vinj < 1$ so long as the frequency is sufficiently large to satisfy the resonance condition in \eqref{eq:rescon}. The same is true for cntr-CAEs, with the caveat that $\krat$ must be sufficiently large as well ($\krat \approx 1$ usually sufficient).  
%Keep in mind also that the aforementioned lower bounds are only the bare minimum $\vinj$ required for $\gamma > 0$ in each case. Realistically, the growth rate does not become appreciable unless $\vinj$ is even somewhat larger. 
These considerations can explain both simulation results and experimental observations. In \HYM simulations of NSTX for a given set of plasma profiles,\cite{Lestz2018APS,Lestz2019sim} co-CAEs are found to require $\vinj \gtrsim 4.5$, whereas cntr-GAEs are excited for a wider range of $\vinj$. In NSTX experiments, counter-propagating modes were more commonly observed than co-CAEs, with the latter appearing only very rarely in NSTX-U experiments which typically operated at much lower $\vinj \lesssim 2$ due to the increased toroidal field strength relative to NSTX. 

A similar comparison can be made for co-GAEs, using the same mode parameters of $\omeganorm = 0.5$ and $\krat = 1$, shown in \figref{fig:wideslowGAE}. Due to the difference in dispersion relation, the co-GAE can sustain a resonant interaction with a fast ion distribution with smaller $\vinj$ than the co-CAE can. The peak growth rate for co-GAEs with these parameters is reduced by an order of magnitude relative to the co-CAE, as expected based on the factor $\omegabar^2\alpha^4$ in front of its FLR function in \eqref{eq:Jlmsmallappx-gae}. Although the co-GAE growth rate peaks at lower $\vinj$ in this example, even at its absolute peak, the co-CAE growth rate is larger. This may explain why co-GAEs driven via the Landau resonance were not observed in NSTX experiments. Furthermore, such modes would have been even more difficult to excite in \HYM simulations, as their drive is strongly enhanced by coupling to the compressional mode, and this coupling is under-estimated in the \HYM model due to the absence of thermal plasma two-fluid effects (see \citeref{Belova2017POP} for a detailed description of the simulation model). 

\subsubsection{Large FLR regime \texorpdfstring{$(\zp \gg 1)$}{}}
\label{sec:fast}

The complementary limit, of large FLR effects, or rapidly oscillating integrand regime due to $\zp \gg 1$ can also be explored. Based on the most unstable modes found in the \HYM simulations, this is not the most common regime for NSTX-like plasmas, but it can occur and is treated for completeness and comparison to the slowly oscillating (small FLR) results. 

This approximation allows the use of the asymptotic form of the Bessel functions: $J_n(\flr) \like \sqrt{2/\pi \flr}\cos\left(\flr - (2n+1)\pi/4\right) \plusord{\flr^{-3/2}}$, which is very accurate for $\flr > 2$. Note also that $\zp \gg 1$ implies $\alpha \ll 1$ since $\zp =\omegabar/\alpha < 1/\alpha$. For both CAEs and GAEs, the FLR function has asymptotic behavior $\J{0}{m}(\flr) \like J_0^2(\flr) \like (1 - \sin(2\flr))/\flr$, where the rapidly varying $\sin(2\flr)$ component will average out in the integrand by the Riemann-Lebesgue Lemma\cite{BenderOrszagStationaryPhase} (see part 1\cite{Lestz2019p1} for further description of this procedure). Then the simplified integral to consider is 

\begin{multline}
\label{eq:dampgamfast}
\gamma \appropto \int_0^{1-\bres} \frac{x^{3/2}(x-\xinj)}{(1-x)^{3/2}}dx \\ 
- \frac{3\dx^2}{4}\int_0^{1-\bres}\frac{\sqrt{x}}{(1-x)^{3/2}}\frac{dx}{1 + \left(\frac{1-x}{4\bres}\right)^{3/2}} \\ 
- \frac{\dx^2}{2}\sqrt{\bres^{-1}-1}e^{-(1-\bres-\xinj)^2/\dx^2}
\end{multline}

Following the same method as in the small FLR regime, first find the marginal stability condition $\gamma = 0$ while neglecting the third term: 

\begin{align}
\xinj &= \frac{h_0(\bres) + \dx^2 h_1(\bres)}{h_2(\bres)} \\ 
h_0(\bres) &= \sqrt{\bres^{-1} - 1}(8+\bres(9-2\bres)) - 15\arccos\sqrt{\bres} \\ 
h_1(\bres) &= 6(-\sqrt{\bres^{-1}-1}+\arccos\sqrt{\bres}) \\
h_2(\bres) &= 4\left(\sqrt{\bres^{-1}-1}(2+\bres)-3\arccos\sqrt{\bres}\right) \\
\Rightarrow \xinj &\approx 1-\bres^{5/7} - \frac{8}{9}\frac{\dx^2}{1 - \bres^{5/7}} \\
\Rightarrow v_0 &= \frac{\vpres}{\left[1 - \frac{1}{2}\left(\xinj + \sqrt{\xinj^2 + \frac{32\dx^2}{9}}\right)\right]^{7/10}}
\label{eq:dampfastvcrit}
%\xinj &= \frac{\sqrt{\bres^{-1} - 1}(8+\bres(9-2\bres)) - 15\arccos\sqrt{\bres} + 6\dx^2(-\sqrt{\bres^{-1}-1}+\arccos\sqrt{\bres})}
%{4\left[\sqrt{\bres^{-1}-1}(2+\bres)-3\arccos\sqrt{\bres}\right]} 
\end{align}

The first part of the approximation ($h_0(\bres)/h_2(\bres)$) is accurate to within $4\%$, while the second part ($h_1(\bres)/h_2(\bres)$) has a maximum relative error of $15\%$, with the error reducing to less than $5\%$ for $\bres > 0.15$. Comparing \eqref{eq:dampfastvcrit} to the analogous instability condition for the same resonance when $\zp \ll 1$, when $\dx = 0$, the $\zp \gg 1$ condition is somewhat more restrictive due to the different exponents, and for finite $\dx$, the correction due to the slowing down part of the $\partial \fb/\partial v$ term is also larger than it is when $\zp \ll 1$, as in \secref{sec:slow}.  

The contribution from the third term in \eqref{eq:dampgamfast} will be treated in the same fashion as in the previous section. Hence, consider solving \eqref{eq:dampgamfast} for marginal stability setting $\xinj = 0$ and approximating $\exp(-x^2) \approx 1/(1 + x^2)$. Then $\dx^2$ can be isolated from a quadratic formula, giving 

\begin{subequations}
\begin{align}
\dx^2 &= \frac{-B - \sqrt{B^2 - 4AC}}{2A} \\ 
\text{where } A &= -\left[h_1(\bres)/4 + 2\sqrt{\bres^{-1} - 1}\right]/4 \\ 
B &= -\left[h_0(\bres) + (1-\bres)^2 h_1(\bres)\right]/4 \\ 
C &= -(1-\bres)^2 h_0(\bres)/4
\end{align}
\end{subequations}

Approximating and inverting this expression gives the following condition for marginal stability at $\xinj = 0$, accurate to within $15\%$

\begin{align}
\bres &\approx ( 1- \dx^{5/6} )^{7/5}
\end{align}

This can be combined with \eqref{eq:dampfastvcrit} to determine the modification to the marginal stability condition required to match the solution at $\xinj = 0$

\begin{align}
v_0 &= \frac{\vpres}{\left[1 - \frac{1}{2}\left(\xinj + \sqrt{\xinj^2 + \frac{32\dx^2}{9}}\right)\right]^{7/10}}
\left(\frac{1 - \dx\sqrt{8/9}}{1- \dx^{5/6}}\right)^{7/10}
\label{eq:widefastCAEgamtail}
\end{align}

\begin{figure}[tb]
\subfloat[\label{fig:widefastCAE}]{\includegraphics[width = \halfwidth]{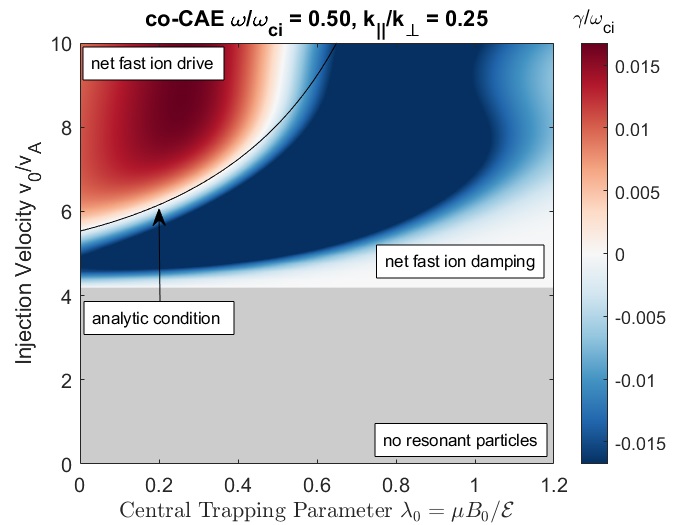}} \\
\subfloat[\label{fig:widefastGAE}]{\includegraphics[width = \halfwidth]{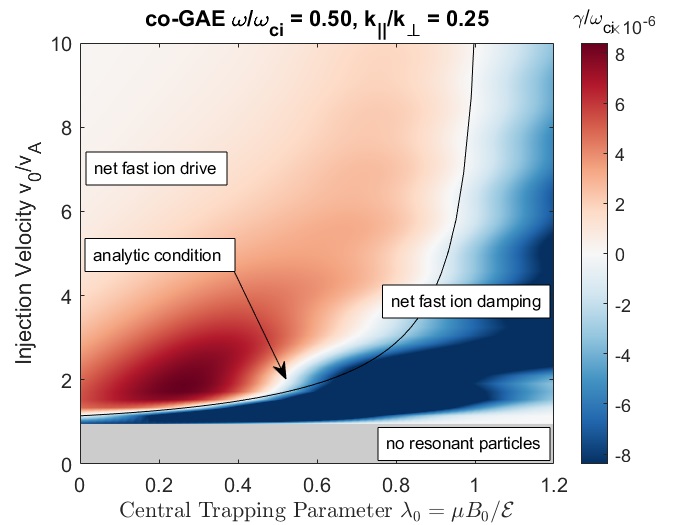}} \\
\caption{Numerical integration of full growth rate expression \eqref{eq:gammabeam} as a function of fast ion distribution parameters $\vinj$ and $\linj$ with $\dx = 0.30$ for a Landau resonance-driven (a) co-CAE and (b) co-GAE in the large FLR regime ($\zp \gg 1$): $\omeganorm = 0.5$ and $\krat = 0.25$, implying $\zp = 2$. Red indicates net fast ion drive, blue indicates net fast ion damping, and gray indicates beam parameters with insufficient energy to satisfy the resonance condition. Solid curve shows approximate stability condition including damping from the tail, derived in \eqref{eq:widefastCAEgamtail}.}
\label{fig:widefastfig}
\end{figure} 

% BEGIN TABLE 
% table to summarize GAE results (l = +/- 1)
\newcommand{\vroom}{\vphantom{{\Huge text}}}
\newcommand{\hroom}{\hspace{1ex}}
\newcommand{\chead}[1]{\multicolumn{1}{c}{#1}}
\begin{table*}\centering
\renewcommand\arraystretch{1.5}
% table to summarize CAE/GAE results (l = 0)
\begin{tabular}{l l l}
\multicolumn{2}{c}{CAE/GAE fast ion drive conditions (Landau resonance)} \\ \hline\hline 
 & \chead{} 
 & \chead{}\vspace{-3ex} \\ 
%\hline
$\zp \lesssim 2$ \hroom\hroom & $v_0 > \dfrac{\vpres}{\left[1 - \frac{1}{2}\left(\xinj + \sqrt{\xinj^2 + 8\dx^2/3}\right)\right]^{5/8}} \left(\dfrac{1 - \dx\sqrt{2/3}}{1-\dx^{4/5}}\right)^{5/8}$ \hroom\vroom \\
$\zp \gg 2$ & $v_0 > \dfrac{\vpres}{\left[1 - \frac{1}{2}\left(\xinj + \sqrt{\xinj^2 + 32\dx^2/9}\right)\right]^{7/10}} \left(\dfrac{1 - \dx\sqrt{8/9}}{1-\dx^{5/6}}\right)^{7/10}$ \vroom \vspace{1ex}\\ 
%$\zp \lesssim 2$ \hroom & $v_0 < \dfrac{\vpres}{(1 - \xinj)^{3/4}}$ \hroom &  $v_0 > \dfrac{\vpres}{(1 - \xinj)^{5/8}}$ \hroom & $v_0 > \dfrac{\vpres}{(1 - \xinj)^{3/4}}$ \hroom\vroom \\
%$\zp \gg 2$ & $v_0 < \dfrac{\vpres}{(1 - \xinj)^{5/6}}$ & $v_0 > \dfrac{\vpres}{(1 - \xinj)^{7/10}}$ & $v_0 > \dfrac{\vpres}{(1 - \xinj)^{5/6}}$ \vroom \vspace{1ex}\\ 
\hline\hline
\end{tabular}

\caption{Approximate net fast ion drive conditions for GAEs and CAEs driven by the Landau resonance in the wide beam approximation, valid for $0.2 < \dx < 0.8$. The quantity $\zp = \kperp\vpres/\omegaci$ is the ``modulation parameter" (see \eqref{eq:zp}) and $x_0 = \lambda_0\omegacires = \vperpz^2/v_0^2$.}
\label{tab:appxcons}
\end{table*}
% END TABLE

This $\zp \gg 1$ marginal stability bound is compared to the numerically evaluated fast ion drive/damping in \figref{fig:widefastfig} for a co-CAE and co-GAE with $\omeganorm = 0.5$ and $\krat = 0.25$ such that $\zp = 2$. While $\zp = 2$ is only marginally within the $\zp \gg 1$ regime, the agreement is still acceptable. Note that in the figures, a maximum value of $\vinj = 10$ is shown, which far exceeds the NSTX range of $\vinj < 6$. This is because the CAE dispersion combined with the resonance condition yields $\zp \approx \omegabar\vpres/\va$ for $\zp \gg 1$, which can not be very large for $\vinj < 6$ considering $\vpres \like v_0/2$ is common, as is $\omeganorm \like 0.5$. The case is different for GAEs since their dispersion yields a parallel resonant velocity that is independent of $\alpha$, such that $\zp$ can be made arbitrarily large by choosing $\alpha$ sufficiently small without constraining the size of $\vpres/\va$. This explains why the co-CAE in the figure has no wave particle interaction when $\vinj < 4$, while an interaction with the co-GAE becomes possible near $\vinj \approx 1$. Although the co-GAE can in principle be driven by fast ions for more accessible values of $\vinj$, note that the growth rate is vastly reduced due to the factor of $\krat^4 \lll 1$. Thus, one would expect that the minuscule magnitude of fast ion drive for the co-GAE shown in \figref{fig:widefastGAE} would be far outweighed by damping on the background plasma. For these reasons, the $\zp \gg 1$ regime is less relevant to modern experimental conditions than the $\zp \ll 1$ regime, except possibly for CAEs with $\omega > \omegaci$ which can be excited at more reasonable values of $\vinj$ (to be addressed in a future work). 

\subsection{Summary of Necessary Conditions for Net Fast Ion Drive}

Here, we briefly summarize the different stability boundaries derived up to this point, along with their ranges of validity. When $1 - \vpres^2/v_0^2 \leq \linj\omegacires$ is satisfied, Landau resonance-driven co-propagating CAEs/GAEs will be net damped by fast ions. All other results address the scenarios when this inequality is not satisfied. When $\dl$ is sufficiently small $(\dl \lesssim 0.10)$, the narrow beam approximation can be made, which yields \eqref{eq:gammasimp}, where the sign of the growth rate depends on $\xinj$ and can be evaluated without further integration. When $\dl$ is sufficiently large $(0.20 \lesssim \dl \lesssim 0.80)$, the wide beam approximation is justified. This includes the nominal NSTX case of $\dl \approx 0.30$. For most of the unstable modes in \HYM simulations, $\zp \lesssim 2$ is also valid, which enables the results contained in the case of a wide beam and slowly oscillating integrand. The complementary limit of $\zp \gg 2$ is also tractable when the beam is sufficiently wide, though this is not the typical case for CAEs and GAEs interacting with fast ions through the Landau resonance. All conditions for the cases involving wide beams are organized in \tabref{tab:appxcons}. 

\section{Preferential excitation as a function of mode parameters} 
\label{sec:stabao}

For fixed beam parameters, the theory can determine which parts of the spectrum may be excited -- complementary to the previous figures which addressed how the excitation conditions depend on the two beam parameters for given mode properties. Such an examination can also illustrate the importance of coupling between the compressional and shear branches due to finite frequency effects on the most unstable parts of the spectra. All fast ion distributions in this section will be assumed to have $\dl = 0.3$ and $\omegacires = 0.9$ for the resonant ions. For the modes driven by the Landau resonance studied here in the small FLR regime, the instability conditions can be written generally as 

\newcommand{\cterm}{d}
\begin{align*}
\cterm^2 &> \vpres^2(\omeganorm,\krat) \numberthis\label{eq:cvpres} \\ 
\cterm &= \frac{v_0}{\va}\left[1 - \frac{1}{2}\left(\xinj + \sqrt{\xinj^2 + \frac{8\dx^2}{3}}\right)\right]^{5/8} \label{eq:dvalue}\\
&\qquad\qquad\times\left(\frac{1-\dx^{4/5}}{1 - \dx\sqrt{2/3}}\right)^{5/8} \numberthis
\end{align*}

In the large FLR regime, $\cterm$ can be replaced by the analogous quantity from \eqref{eq:widefastCAEgamtail}. Determining the unstable regions of the spectrum as a function of $\omeganorm$ and $\krat$ therefore relies on the dependence of $\vpres$ on these quantities. This dependence can be well approximated as 

\begin{align}
\vpres^{CAE} &\approx \sqrt{\frac{1}{\alpha^2} + 1 + \omegabar} \label{eq:vprescae}\\ 
\vpres^{GAE} &\approx \sqrt{1 - \omegabar^{\frac{2 + \alpha^2}{1 + \alpha^2}}} \label{eq:vpresgae}
\end{align}

These expressions have a maximum relative error of $3\%$ and $6\%$ respectively for $0 < \omeganorm < 1$ and all values of $\krat$. Using these expressions, the approximate stability conditions become 

\begin{align}
0 &< \left(\frac{\omega}{\omegaci}\right)^{CAE} < \cterm^2 - \left(1 + \frac{1}{\alpha^2}\right) \label{eq:caezao}\\ 
\left(1 - \cterm^2\right)^{\frac{1 + \alpha^2}{2 + \alpha^2}} &< \left(\frac{\omega}{\omegaci}\right)^{GAE} \label{eq:gaezao} < 1
\end{align}

\newcommand{\thirdwidth}{.35\textwidth}
\begin{figure*}[tb]
\hspace*{-2ex}
\subfloat[\label{fig:caezao30}]{\includegraphics[width = \thirdwidth]{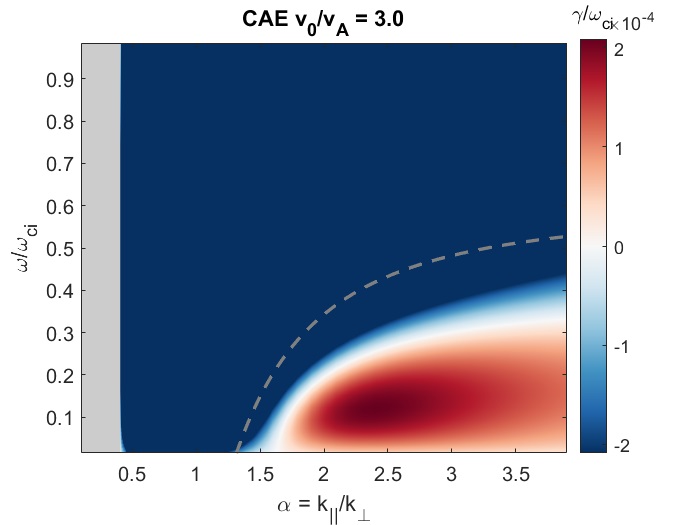}}
\hspace*{-3ex}
\subfloat[\label{fig:caezao35}]{\includegraphics[width = \thirdwidth]{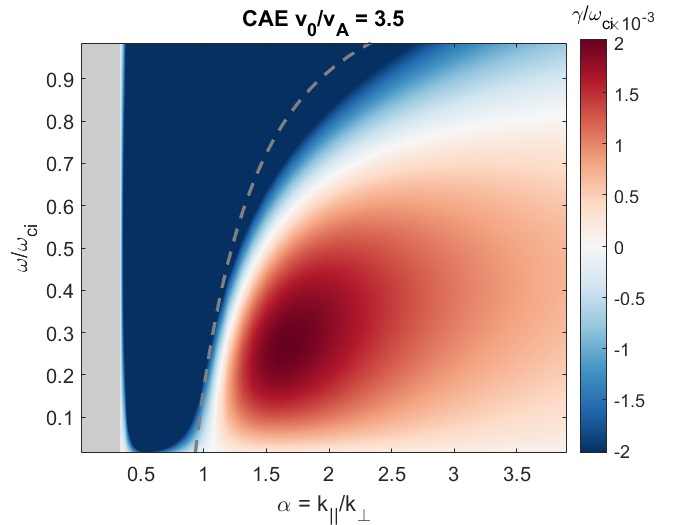}}
\hspace*{-3ex}
\subfloat[\label{fig:caezao40}]{\includegraphics[width = \thirdwidth]{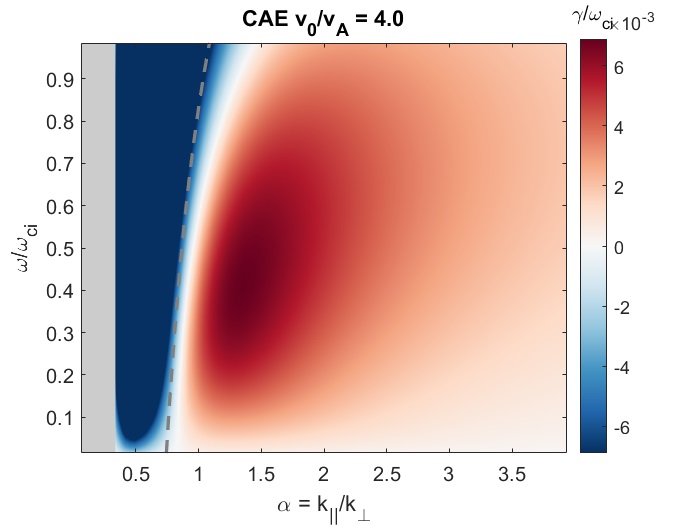}} \\ 
\hspace*{-2ex}
\subfloat[\label{fig:gaezao15}]{\includegraphics[width = \thirdwidth]{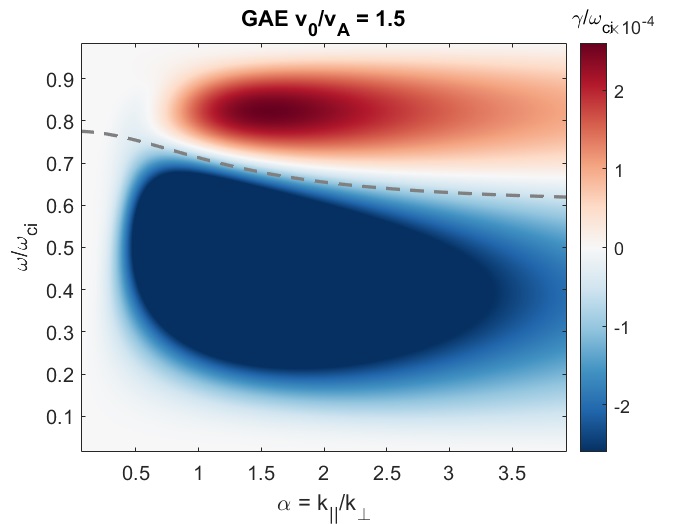}}
\hspace*{-3ex}
\subfloat[\label{fig:gaezao20}]{\includegraphics[width = \thirdwidth]{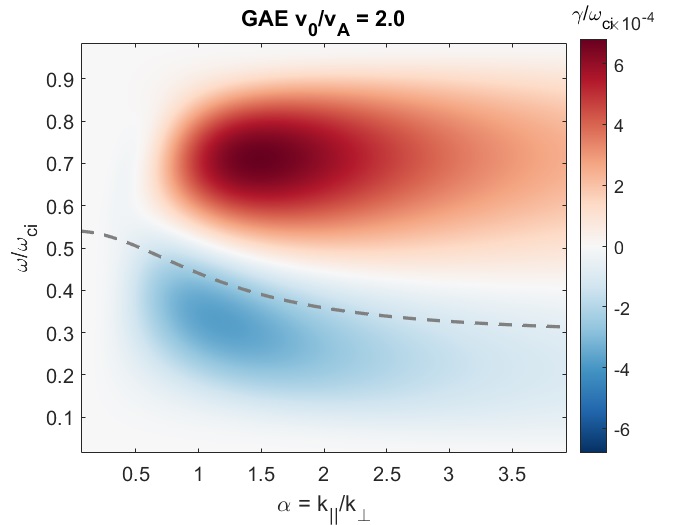}}
\hspace*{-3ex}
\subfloat[\label{fig:gaezao25}]{\includegraphics[width = \thirdwidth]{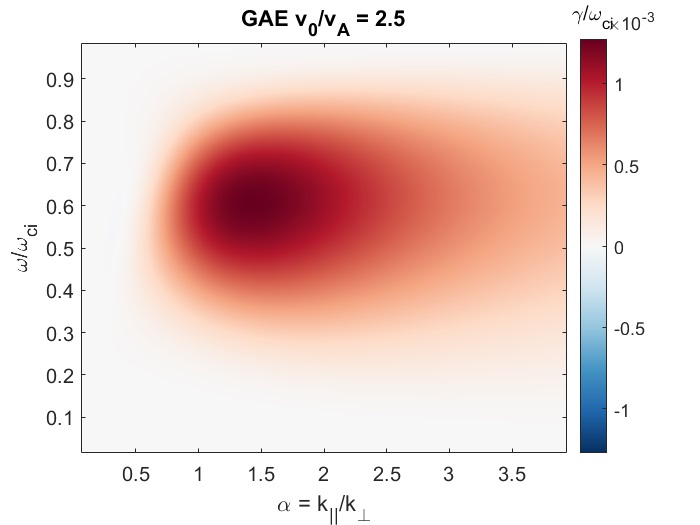}}
\caption{Numerically calculated fast ion drive/damping for Landau resonance-driven (a - c) co-CAEs for $\vinj = 3.0 - 4.0$ and (d - f) co-GAEs for $\vinj = 1.5 - 2.5$ as a function of $\omegabar = \omeganorm$ and $\alpha = \krat$, when driven by a beam distribution with $\linj = 0.7$, $\dl = 0.3$, and assuming $\omegacires \approx 0.9$. Red corresponds to net fast ion drive, blue to damping, and gray to regions excluded by the resonance condition. Gray curves indicate approximate marginal stability conditions.}
\label{fig:bothzao}
\end{figure*}

A comparison between these boundaries and the numerically integrated expression for growth rate is shown in \figref{fig:bothzao}. There, a fast ion distribution with $\linj = 0.7$ is assumed, similar to NSTX conditions, and the calculation is shown for different values of $\vinj$. Consider first the case of the CAEs. Note that there is a minimum value of $\alpha$ below which all frequencies are stable. This follows from \eqref{eq:caezao} when $\cterm^2 < 1 + 1/\alpha^2$. For small values of $\vinj$, only small values of $\omeganorm$ can be driven by the fast ions, even though the resonance condition is satisfied for all frequencies. For larger values of $\vinj$, the frequency dependence of this boundary becomes very weak, with the boundary converging simply to $\alpha > \alphamin$. Note that if coupling to the shear mode were neglected, $\vpres$ for the CAEs would be independent of $\alpha$, which would remove the frequency dependence of the marginal stability boundary even in the case of small $\vinj$. The dashed gray curves plot \eqref{eq:caezao}, demonstrating qualitative agreement with the numerically evaluated expression. The quantitative disagreement is mostly inherited from the inaccuracy of the \emph{ad-hoc} correction for the damping coming from the tail of the distribution, which used a factor to match the solution at $\linj = 0$, leading to larger errors at larger $\linj$ such as $\linj = 0.7$ used for these plots. 

Considering now the GAEs, not only is their drive only made possible due to coupling to the compressional branch, as discussed in \secref{sec:derivation}, but the unstable spectrum can also only be described when considering the coupled dispersion relation. Suppose instead that the simplified dispersion were used. Then $\vpres \approx 1$ would be true for the GAEs, implying $\cterm^2 > 1$ for instability, which is completely independent of $\omeganorm$ and $\krat$. However, \figref{fig:bothzao} clearly shows a minimum frequency for instability when $\vinj$ is not too large. This results from coupling to the compressional branch, which results in the modification to $\vpres$ included in \eqref{eq:vpresgae}. The dashed curves on \figref{fig:bothzao} compare the approximate instability conditions to the numerically integrated growth rate, showing that this correction is qualitatively captured. Again, there is some quantitative mismatch between the analytic condition and the true marginal stability boundary due to the less accurate treatment of damping from the tail. Moreover, it is worth pointing out that unlike the co-CAEs, as $\vinj$ is increased for the co-GAEs, it becomes possible to destabilize modes with \emph{smaller} frequencies. 

Note that for sufficiently large values of $\vinj$ (determined by $\cterm^2 > 1$), the GAEs can be strictly driven for all values of $\omeganorm$ and $\krat$. Such an example is shown in \figref{fig:gaezao25}. However, the drive can become extremely small for regions of this parameter space far from the most favorable parameters, where modes will be stabilized by any damping mechanisms (thermal plasma, continuum) not considered here. The peak growth rate occurs near $\alpha \approx 1.5$ and $\omeganorm \approx 0.6$ in this case. This can be qualitatively understood from the form of the FLR function. For very small $\alpha$, the coefficient $\alpha^4$ in \eqref{eq:Jlmsmallappx-gae} substantially decreases the growth rate. In contrast, at large $\alpha$, the coefficient in front of the Bessel function can be order unity, however the argument $\flr = \zp\sqrt{x/(1-x)}$ becomes small since $\zp = \omegabar/\alpha$, and hence $J_1^2(\omegabar/\alpha) \propto 1/\alpha^2$ for $\alpha \gg 1$. The local maximum in frequency can be understood similarly, as at low frequency, there is a coefficient $\omegabar^2$ in front of the Bessel function, and also the Bessel function will expand as $\omegabar^2$. For the limit of $\omeganorm \rightarrow 1$, the coefficient in \eqref{eq:Jlmsmallappx-gae} vanishes for the GAEs. 

No special weight should be assigned to the values of $\vinj$ used in \figref{fig:bothzao} in relation to the shapes of the stability boundaries in general, since these conditions also depend on $\linj$. They are relevant to NSTX since the value used in the figure, $\linj = 0.7$, is characteristic of the neutral beam geometry used for that experiment. For instance, for a different value of $\linj$, the co-GAEs would become unstable for all frequencies (\eg \figref{fig:gaezao25}) at some other value of $\vinj$. Likewise, the co-CAE boundary will also converge to $\alpha > \alphamin$ for a value of $\vinj$ depending on $\linj$. 

\section{Experimental Comparison} 
\label{sec:expcomp}

Co-CAEs were studied in depth in NSTX in many discharges in \citeref{Fredrickson2013POP} and manually analyzed to determine the toroidal mode number and frequency of each observed eigenmode (in contrast to the database of cntr-GAEs discussed in part 1, which was more massive and therefore relied on spectrum-averaged quantities calculated via automated analysis). Co-CAEs can be unambiguously distinguished\cite{Fredrickson2013POP,Appel2008PPCF,Sharapov2014PP} from cntr-GAEs due to the direction of propagation and the absence of other modes in the high frequency range studied ($\omeganorm \gtrsim 0.5$). From a simplified 2D dispersion solver, these high $\abs{n} (> 10)$ modes were inferred to be localized in a potential well near the low field side edge, typically with low $\abs{m} \lesssim 2$. It is worth noting that these high frequency co-CAEs were mostly observed when a low frequency $n = 1$ kink mode was present, though the source of their nonlinear interaction is not precisely known.\cite{Fredrickson2013POP}

Whereas the cntr-GAE stability condition\cite{Lestz2019p1} yielded lower and upper bounds on the unstable range of frequencies for a given $(\linj,\vinj)$, the marginal stability condition for co-CAEs (given in \eqref{eq:caezao}) instead yields a lower bound on the allowed value of $\krat$ in the low coupling limit of $\omeganorm \ll 1$, which is usually more restrictive than the lower bound on $\krat$ resulting from the requirement $\vpres < v_0$. Hence, one of these lower bounds will always be redundant. An upper bound on $\krat$ can be derived heuristically, considering that the CAEs are trapped in a local effective potential well\cite{Gorelenkova1998POP,Kolesnichenko1998NF,Smith2003POP,Smith2009PPCF,Smith2017PPCF} 
of characteristic width $\Delta R\approx R_0/2$. To satisfy this constraint, an integer number of half wavelengths must fit within the potential well, such that $k_{R,\text{min}} = \pi/\Delta R$. Similarly, poloidal symmetry requires $k_{\theta,\text{min}} = m/a$ for integer $m$. Hence, $\kperpmin \approx (2\pi/R_0)\sqrt{1 + (R_0/2\pi a)^2} \approx 2\pi/R_0$. Moreover, $\kpar \approx k_\phi = n/R_0$ is a reasonable approximation for the observed high $\abs{n}$, low $\abs{m}$ modes. Hence, $\krat_\text{max} = \kpar / \kperpmin \approx n/2\pi$. 

Although $\kperp$ is not a reliably measured experimental quantity, it can be inferred from the measured frequency and toroidal mode number using the approximate dispersion $\omega \approx k\va$, such that $\krat = 1/\sqrt{\omega^2 R_0^2 / n^2 \va^2 - 1}$. Within this local framework, $\va$ is evaluated near the plasma edge, where the mode exists, to calculate $\krat$. 

\begin{figure}[tb]
\includegraphics[width = \halfwidth]{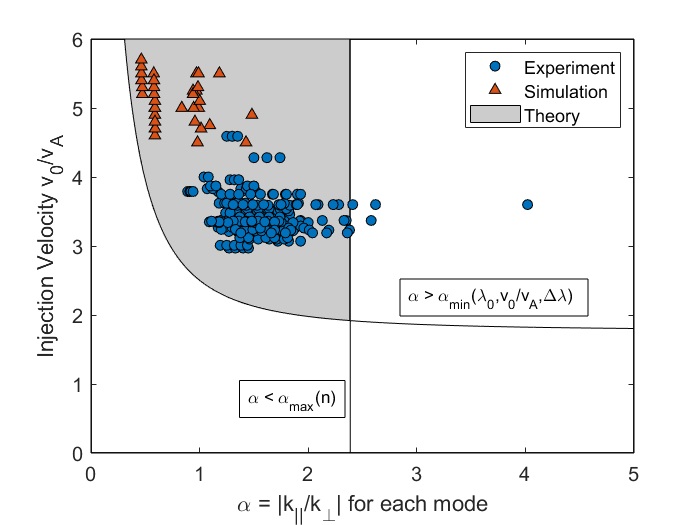}
\caption{Comparison between theory, simulations, and experiment. Blue circles represent individual co-CAE modes from NSTX discharges. Red triangles show co-CAEs excited in \HYM simulations with $\linj = 0.7$. Theory predicts net fast ion drive in the shaded region between the two curves.}
\label{fig:allcomp-co}
\end{figure}

The comparison of these bounds with the experimental observations (blue circles) and simulation results (red triangles) is shown in \figref{fig:allcomp-co}. The curve is calculated from \eqref{eq:caezao} using $\linj = 0.65$, which was the average value for the studied discharges. Also, $N = 1$ was chosen for consistency with the formula used to calculate the wavenumber from the measured frequency. The straight line represents the heuristic upper bound on $\krat$ using $n = 15$, which was the maximum value in the experimental database. Hence, our theory predicts net fast ion drive in the shaded region between the curve and vertical line. Only simulations with $\linj = 0.7$ are shown in order to remain close to the average value of $\linj = 0.65$ for the experimental conditions shown. All points have $\krat > \alphamin$, in agreement with \eqref{eq:caezao}, and most of the points are also consistent with $\krat < \alphamax$. When calculating these boundaries for the specific properties of each mode, it is found that all of the simulation points fall within the allowed range, while $82\%$ of the experimental co-CAEs agree with theory. The outliers with $\krat > \alphamax$ could be due to either a wider potential well width in those discharges or slight errors in the $n$ number identification due to limited toroidal resolution. Such an experimental comparison can not be made with co-GAEs at this time, as none were identified in NSTX, likely due to their reduced growth rate relative to the co-CAEs. 

Further analysis of the linear simulation results shown on \figref{fig:allcomp-co} will be described in detail in a forthcoming paper.\cite{Lestz2019sim} The simulation set up and properties of the modes can be found in \citeref{Lestz2018POP}. The simulations used equilibrium profiles from the well-studied H-mode discharge $\#141398$,\cite{Fredrickson2013POP,Crocker2013NF,Crocker2017NF,Belova2017POP} and fast ion distributions with the same $(\lambda,v)$ dependence studied in this work, and given in \eqref{eq:Fdistr}. The $\pphi$ dependence was fit from \TRANSP to a power law, as described in \citeref{Belova2017POP}. The peak fast ion density in all cases is $n_b/n_e = 5.3\%$, matching its experimental value in the model discharge. 

\section{Summary and Discussion}
\label{sec:summary}

The fast ion drive/damping for compressional (CAE) and global (GAE) \Alfven eigenmodes due to the Landau resonance has been investigated analytically for a model slowing down, beam-like fast ion distribution, such as those generated by neutral beam injection in NSTX. The local growth rate includes contributions to all orders in $\krat$ and $\omeganorm$, addressing parameter regimes that were not treated by previous work studying this instability\cite{Belikov2003POP,Belikov2004POP}. Retaining finite $\omeganorm$ and $\krat$ was demonstrated to be important for capturing the coupling between the shear and compressional branches (present due to two-fluid effects in our model), which was in turn vital to the existence of the co-GAE instability. The full FLR dependence was also kept in this derivation, as in previous work. The dependence of the fast ion drive was studied as a function of four key parameters: the beam injection velocity $\vinj$, the beam injection geometry $\linj = \mu B_0 / W$, the mode frequency $\omeganorm$, and the direction of the wave vector $\krat$. It was shown that CAEs require relatively large $\vinj$ in order to have an appreciable growth rate, explaining why they were observed much less frequently in NSTX-U than NSTX. Moreover, the growth rate of the GAE carries an additional small coefficient of $(\omeganorm)^2\krat^4$ relative to the CAE, suggesting why these are rarely observed. 

Without further approximation, the derived growth rate led to an immediate corollary: when $1 - \vpres^2/v_0^2 \leq \linj\omegacires$, only damping occurs from the Landau resonance. For cases where this condition is not satisfied, approximate conditions for net fast ion drive were derived by making experimentally relevant approximations. Previous analytic conditions\cite{Belikov2003POP,Belikov2004POP} for net fast ion drive of CAEs driven by the Landau resonance were limited to delta functions in $\lambda$, which are a poor approximation for fast ions generated by NBI. In contrast, the instability conditions derived here result from integrating over the full beam-like distribution with finite width in velocity space. It was found in \secref{sec:narrow} that the approximation of a narrow beam was only valid when $\dl \lesssim 0.1$, much smaller than the experimental value of $\dl \approx 0.3$. Consequently, our more general derivation allows for instability at any value of $\kperp\rhob$, whereas prior work concluded a limited range. 

The approximation of a sufficiently wide beam in conjunction with a small or large FLR assumption yielded an integral in the growth rate expression which could be evaluated exactly and led to useful conditions for net fast ion drive, listed in \tabref{tab:appxcons}. In particular, the condition for a wide beam and small FLR effects ($\zp = \kperp\vpres/\omegaci \lesssim 2$) is typically applicable to NSTX conditions, as determined from observations and simulations of these modes. 

Comparison between the numerical integration of the analytic expression for growth rate and the approximate stability boundaries indicates strong agreement within the broad parameter regimes that they apply. Since these stability conditions depend on both fast ion parameters ($\linj, \vinj$) and mode parameters $(\omeganorm,\krat)$, they can provide information both about how a specific mode's stability depends on the properties of the fast ions, as well as the properties of the modes that may be driven unstable by a specific beam distribution. Namely, co-propagating CAEs are unstable for sufficiently large $\krat$, nearly independent of frequency when $\vinj$ is sufficiently large. In contrast, when $\vinj$ is not too large, GAEs can only be excited at high frequencies. The approximate condition for CAE stability was compared against NSTX data from many discharges, yielding greater than $80\%$ agreement, demonstrating the utility of these results in interpreting observations and guiding future experiments. One area of ongoing work is the application of this theory to predict ways to stabilize co-propagating modes with the addition of a second beam source, complementary to the cntr-GAE suppression observed in NSTX-U\cite{Fredrickson2017PRL} with small amounts of power in the new, off-axis beam sources. 

It is worth reminding one final time of the simplifications used in deriving these results. Contributions from the gradient in $\pphi$ were not analyzed, though this is not expected to be a substantial correction based on past simulations\cite{Belova2017POP}. The calculation was also local, not accounting for spatial profiles or mode structures. Consequently, the magnitude of the drive/damping shown in figures should not be considered absolute, but rather relative. Lastly, the net drive conditions do not include sources of damping coming from the background plasma, so they should be interpreted as necessary but not sufficient conditions for instability. Careful analysis of these damping sources and their dependence on all of the parameters studied here (including kinetic contributions from the large fast ion current) is left for future work. 

\section{Acknowledgments}
\label{sec:acknowledgments}

The authors are grateful to E.D. Fredrickson for providing additional experimental data for comparison, as well as A.O. Nelson for fruitful discussions. The simulations reported here were performed with computing resources at the National Energy Research Scientific Computing Center (NERSC). The data required to generate the figures in this paper are archived in the NSTX-U Data Repository ARK at the following address:
\url{http://arks.princeton.edu/ark:/88435/dsp011v53k0334}. This research was supported by the U.S. Department of Energy (NSTX contract DE-AC02-09CH11466 and DE-SC0011810).

\appendix 

\section{Calculations for \secref{sec:slow}}
\label{app:dampref}

Details from the calculations in \secref{sec:slow} are listed here for reference. The full form of \eqref{eq:CAEdampgamsym} is 

\begin{subequations}
\begin{align}
\xinj &= g_0(\bres) -\dx^2 g_1(\bres) \\ 
\text{where } g_0(\bres) &= \frac{1 - 8\bres + 8\bres^3 - \bres^4 - 12\bres^2\log\bres}{1 - 6\bres + 3\bres^2 + 2\bres^3 - 6\bres^2\log\bres} \\ 
\text{and } g_1(\bres) &= \frac{A + B + C + D}{32\left(1 - 6\bres + 3\bres^2 + 2\bres^3 - 6\bres^2\log\bres\right)} \\ 
A &= 12\left(1 + \sqrt{\bres} - 8\bres + 6\bres^2\right) \\ 
B &= 2\sqrt{3}(1 + 8\bres)\arctan\left(\frac{\bres^{-1}-1}{\sqrt{3}}\right) \\ 
% C &= \log3 + \log\left(4 + \bres^{-1} - 2\bres^{-1/2}\right) - 2\log\left(2 + \bres^{-1}\right) \\ 
C &= \log\left[3\left(\frac{(1 - 2\sqrt{b})^2 + 2\sqrt{b}}{(1 + 2\sqrt{b})^2}\right)\right] \\
\nonumber
D &= 2\bres\log\left[\frac{81}{\bres^3(8 + \bres^{3/2})^2}\right] \\ 
&+ \log\left[\frac{1}{3} + \frac{2\sqrt{\bres}}{(1 - 2\sqrt{\bres})^2 + 2\sqrt{\bres}}\right] 
\end{align}
\end{subequations}

The solution of \eqref{eq:bdampcrit} is a quadratic formula for $\dx^2$, given by 

\begin{subequations}
\begin{align}
\dx^2 &= \frac{-B - \sqrt{B^2 - 4AC}}{2A} \\ 
\text{where } A &= I_\text{damp} - (\bres^{-1} - 1)^2/2 \\ 
B &= I_\text{drive} + (1-\bres)^2 I_\text{damp} \\ 
C &= (1-\bres)^2 I_\text{drive} 
\end{align}
\end{subequations}

And the integrals are evaluated as 

\begin{subequations}
\begin{align}
I_\text{drive} &= \int_0^{1-\bres} \frac{x^4}{(1-x)^3}dx \\
&= \frac{1-\bres(8 -\bres^2(8-\bres))}{2\bres^2} - 6\log\bres 
\end{align}

\begin{widetext}
\begin{align}
\numberthis
I_\text{damp} &= - \frac{3}{4}\int_0^{1-\bres} \frac{x^2}{(1-x)^3}\frac{dx}{1 + \left(\frac{1-x}{4\bres}\right)^{3/2}} \\
\nonumber 
&= -\frac{1}{64\bres^2}\left\{
%
% vphantom for spacing 
%
\vphantom{\left. \left. +4\bres\log\left[3\left(\frac{1+(4\bres)^{3/2}}{\bres^{3/2}}\right)^{4\bres}\left(\frac{\left((1-2\sqrt{\bres})^2+2\sqrt{\bres}\right)\left(1+2\sqrt{\bres}\right)^2}{\bres^2}\right)\right] \right] + \log(3 - 6\sqrt{\bres}+12\bres)\right\}}
%
% resume math 
%
2\sqrt{3}(1 + 8\bres)\arctan\left(\frac{-1 + \bres^{-1/2}}{\sqrt{3}}\right) \right.
 -2\left[
%
% vphantom for spacing  
%
 \vphantom{\left. \left. +4\bres\log\left[3\left(\frac{1+(4\bres)^{3/2}}{\bres^{3/2}}\right)^{4\bres}\left(\frac{\left((1-2\sqrt{\bres})^2+2\sqrt{\bres}\right)\left(1+2\sqrt{\bres}\right)^2}{\bres^2}\right)\right] \right] + \log(3 - 6\sqrt{\bres}+12\bres)\right\}}
%
% resume math 
%
 6(-1-\sqrt{\bres}+8\bres)-4\bres^2(9+8\log{3}) + \log(1 + 2\sqrt{\bres}) \right. \\
&\left. \left. +4\bres\log\left\{3\left(\frac{1+(4\bres)^{3/2}}{\bres^{3/2}}\right)^{4\bres}\left(\frac{\left((1-2\sqrt{\bres})^2+2\sqrt{\bres}\right)\left(1+2\sqrt{\bres}\right)^2}{\bres^2}\right)\right\} \right] + \log(3 - 6\sqrt{\bres}+12\bres)\right\}
\end{align}
\end{widetext}
\end{subequations}

%
%\bibliography{../all_bib} 
\bibliography{all_bib} 
\end{document}